\documentclass[11pt]{article}
\usepackage{hyperref}
\pdfoutput=1
\begin{document}
\title{How Capillary Rafts Sink}

\author{S. Protiere$^{1}$, M. Abkarian$^{2}$, J. Aristoff$^{3}$ and H. Stone$^{3}$ \\
\\\vspace{6pt} $^{1}$CNRS-Institut Jean le Rond d'Alembert, Paris, France \\
$^{2}$CNRS-LCVN, Montpellier, France \\
$^{3}$MAE-Princeton University, Princeton, USA\\}

\maketitle

\begin{abstract}
We present a fluid dynamics video showing how capillary rafts sink. Small objects trapped at an interface are very common in Nature (insects walking on water, ant rafts, bubbles or pollen at the water-air interface, membranes...) and are found in many multiphase industrial processes. Thanks to Archimedes principle we can easily predict whether an object sinks or floats. But what happens when several small particles are placed at an interface between two fluids. In this case surface tension also plays an important role. These particles self-assemble by capillarity and thus form what we call a "capillary raft". We show how such capillary rafts sink for varying sizes of particles and define how this parameter affects the sinking process.
\end{abstract}
\section{Submission Content}
There are two fluid dynamics videos contained in this submission for the Gallery of Fluid Motion of the APS DFD Meeting 2010.

\begin{enumerate}
\item Protiere-GoFM2010Hi.mov - high quality video for the DFD2010 Gallery of Fluid Motion display.
\item Protiere-GoFM2010Lo.mov - low quality video suitable for web viewing.

\end{enumerate}
\end{document}